\documentclass[11pt,twoside]{article}
\usepackage{asp2010}

\resetcounters

\begin{document}
\label{jonaycontribution}

%
\title{Volatile and refractory abundances of solar analogs with 
planets}
\author{J.~I.~Gonz\'alez~Hern\'andez,$^{1,2}$
E.~Delgado-Mena,$^{3}$ S.~G.~Sousa,$^{3}$, G.~Israelian,$^{1,2}$ 
N.~C.~Santos,$^{3,4}$ and S.~Udry$^{5}$	  
\affil{
$^1$Instituto de Astrof{\'\i }sica de Canarias (IAC), 
E-38205 La Laguna, Tenerife, Spain: jonay@iac.es\\
$^2$Depto. Astrof{\'\i }sica, Universidad de La 
Laguna (ULL), E-38206 La Laguna, Tenerife, Spain\\
$^3$Centro de Astrof\'isica, Universidade do Porto, Rua
das Estrelas, 4150-762 Porto, Portugal \\
$^4$Departamento de F\'{\i}sica e Astronomia, Faculdade 
de Ci\^encias, Universidade do Porto, Portugal\\
$^5$Observatoire Astronomique de l'Universit\'e de
Gen\`eve, 51 Ch. des Maillettes, -Sauverny- Ch1290, Versoix, 
Switzerland
}
}


\begin{abstract}

We present a detailed abundance analysis of high-quality 
HARPS, UVES and UES spectra of 95 solar analogs, 33 with and 
62 without detected planets. These 
spectra have S/N~$> 350$. We investigate the possibility that 
the possible presence of terrestrial planets could affect 
the volatile-to-refratory abundance ratios.

We do not see clear differences between stars with and 
without planets, either in the only seven solar twins or 
even when considering the whole sample of 95 solar analogs 
in the metallicity range $-0.3<\mathrm{[Fe/H]}<0.5$.
We demonstrate that after removing the Galactic 
chemical evolution effects the possible differences 
between stars with and without planets in these samples 
practically disappear and the volatile-to-refractory 
abundance ratios are very similar to solar values.    

We investigate the abundance ratios of volatile and refractory 
elements versus the condensation temperature of this sample 
of solar analogs, in particular, paying a special attention to 
those stars harbouring super-Earth-like planets.

\end{abstract}

\section{Introduction\label{secintro}}

The last 10 years of spectroscopic observations dedicated to the
search of extrasolar planets have provided an unprecedented set of
high-quality spectra. Indeed, the HARPS GTO 
program~\citep{may03} have produced a spectroscopic database 
containing a substantial amount of high-resolution and 
high signal-to-noise spectra of G-type stars. 
These high-quality data have been used to produce detailed 
abundance analysis \citep{nev09,adi12a,adi12b}. In particular,
a subsample of very high-quality spectra of solar analogs were 
analyzed to derive very accurate abundances of
24 chemical elements~\citep{gon10}. 
Previously, \citet{mel09} noticed that the abundance pattern 
of the Sun reveals a deficiency in refractory elements with 
respect to the volatile content 
when comparing with the mean abundance pattern of 11 solar twins. 
They found a clear decreasing trend of the mean abundance differences, 
$\Delta {\rm [X/Fe]_{\rm SUN-STARS}}$, versus the condensation
temperature, $T_C$, and suggested that this issue was connected
with the presence of terrestrial planets in the solar planetary 
system.
\citet{ram09} found a similar behaviour than that of \citet{mel09} 
when considering a larger sample of solar analogs. 

Later, \citet{gon10} studied in detail the chemical abundances of 
a subsample of solar analogs in the HARPS 
program with and without planets with very high-quality HARPS, 
UVES and UES spectroscopic data. 
These authors demonstrated that the abundance ratios, [X/Fe], of 
solar analogs with and without planets exhibit very similar 
Galactic chemical evolution trends. They also found a decreasing 
trend of the 
mean abundance differences, $\Delta {\rm [X/Fe]_{\rm SUN-STARS}}$, 
versus the condensation temperature, $T_C$, of the small subsample 
of seven solar twins in the HARPS data, two with and five without 
known planets. However, despite the higher quality of the HARPS data,
the scatter of the data points around the linear fits was larger than
that of the results presented by \citet{mel09}. 
Also, the mean abundance differences of all G-type dwarfs with and 
without planets versus the condensation
temperature also present similar behaviours. 

The mean abundance trends of their sample of solar analogs with 
and without planets get null when taking into account the impact of 
the Galactic trends \citep{gon11sea}. In addition, 
two stars hosting super-Earth-like planets, HD~1461 and HD~160691,
also show practically zero values of volatile-to-refractory
abundance ratios compared to solar when subtracting the effects 
of the Galactic chemical evolution \citep{gon11iau}. 
Here, we revisit the chemical analysis of solar analogs with and 
without planets taking into account the discovery of new low-mass
exoplanets~\cite[e.g.][]{may11}.   

\section{Observations}

The spectroscopic data used in this work were taken with the 
HARPS, UVES, and UES spectrographs at the 3.6-m ESO, 8-m VLT, 
and 4.2-m WHT telescopes, installed in the La Silla Paranal 
Observatory, ESO (Chile), 
and in the Spanish Observatorio del Roque de los Muchachos of 
the Instituto de Astrof{\'\i}sica de Canarias, in the island 
of La Palma, respectively. The selected spectra of solar analogs
were observed with HARPS and UVES spectrographs at resolving power 
of $R\sim 110,000$ and for some UVES and UES spectra, at 
$R\sim 85,000$ and $\sim 65,000$, respectively. 
All the spectra used in this work 
have S/N~$> 350$. These very high-quality data have on average a 
S/N ratio of roughly 850. 

\section{Stellar parameters and abundance analysis}

The stellar parameters and metallicities of the whole sample 
of stars were computed by applying the equivalent width method, 
described in \citet{sou08}, to a set of FeI-II lines 
measured with the code ARES \citep{sou07}, 
and evaluating the excitation and ionization equilibria. 
The chemical abundance derived for each spectral line was
computed using the 2002 version of the LTE code MOOG \citep{sne73}, 
and a grid of Kurucz ATLAS9 plane-parallel model atmospheres
\citep{kur93}. 

We determine the mean abundance of each element relative to its 
solar abundance by computing the line-by-line mean difference.
Our fully differential analysis is, at least, internally consistent.
We use the HARPS spectrum of \emph{Ganymede}, a Jupiter's satellite,
as solar reference, which has a 
S/N~$\sim400$~\citep[see][for further details]{gon10}.

\section{Discussion}

\subsection{Solar analogs with and without planets}

\begin{figure}[!ht]
\centering
\includegraphics[width=4.7cm,angle=90]{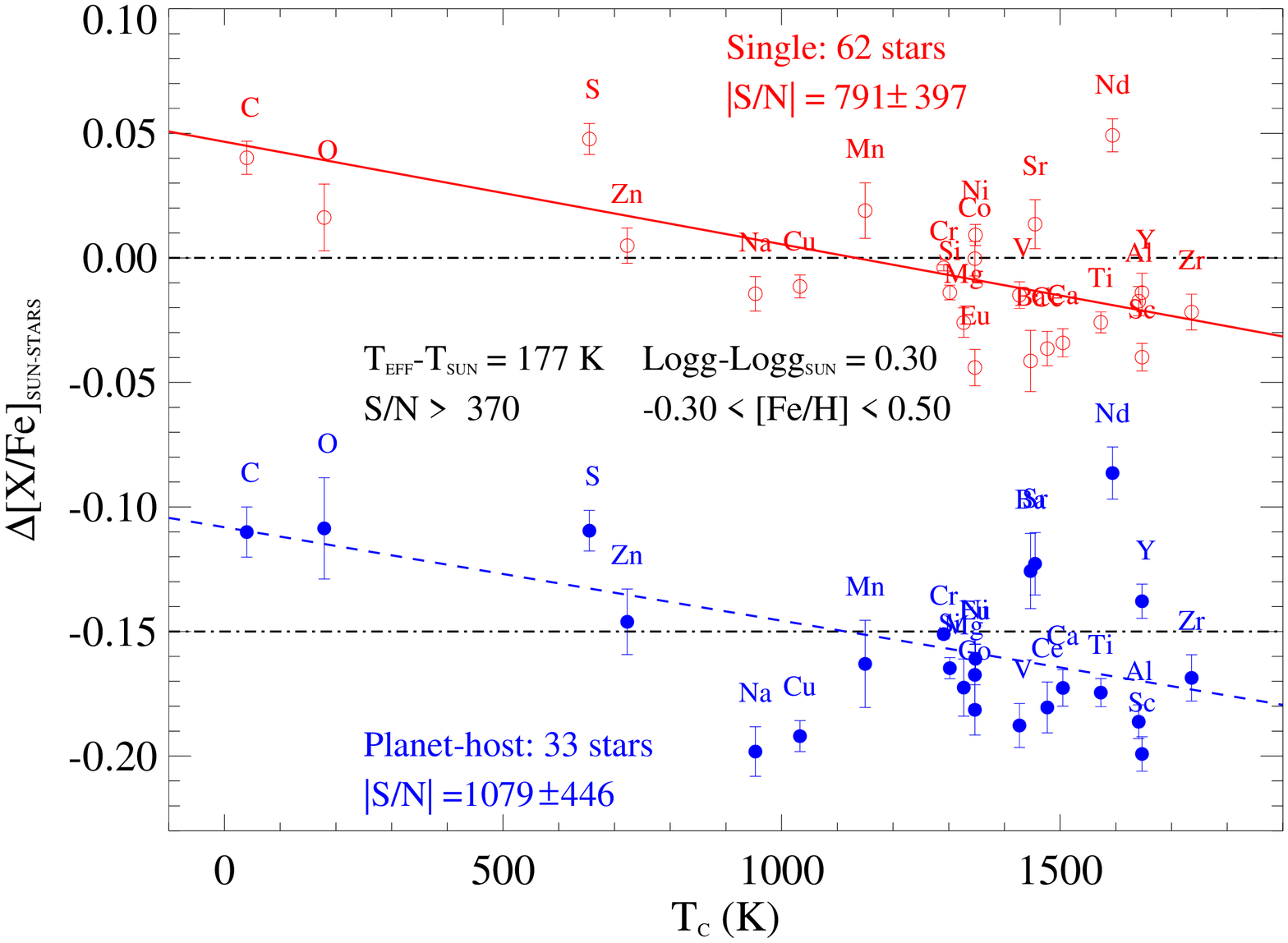}
\includegraphics[width=4.7cm,angle=90]{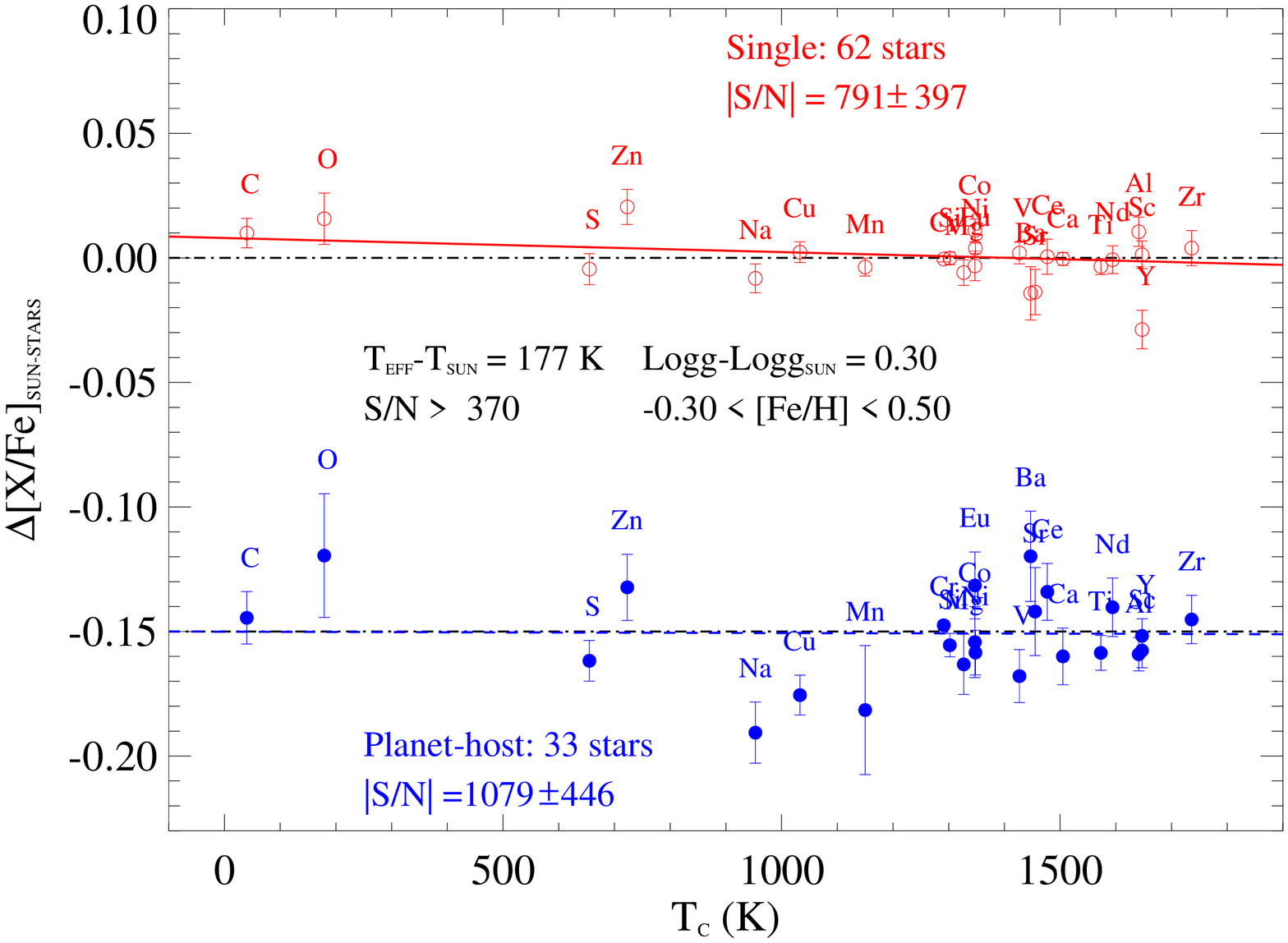}
\caption{\scriptsize{{\it Left panel:} Mean abundance differences, 
$\Delta {\rm [X/Fe]_{SUN-STARS}}$, between the Sun, and 
33 planet-host stars (blue filled circles) and 62 ``single''
stars (red open circles) of the whole sample 
of solar analogs. Error bars are the standard deviation from the
mean divided by the square root of the number of stars. 
Linear fits to the data points weighted with the error bars 
are also displayed for planet hosts (blue dashed line) and 
``single'' stars (red solid line).
An arbitrary shift of -0.15~dex has been applied to the abundances 
of the planet-host stars, for the sake of clarity.
{\it Right panel:} Same as left panel but after
correcting each element abundance ratio of each star using a
linear fit to the Galactic chemical trend of the corresponding
element at the metallicity of each star.}}   
\label{fsa}
\end{figure}

The sample of solar analogs have 62 stars without known planets and 
33 stars with planets. 
The mean abundance ratios of the whole sample
are displayed in Fig.~\ref{fsa}, and they seem to
behave in a similar way for both stars with and without planets. 
These abundance differences exhibit a decreasing trend towards
increasing condensation temperature although some scatter around the
linear fits to the data points is appreciable. We consider that the
Galactic chemical effects have an impact on these results even if the
metallicity range is relatively narrow. Thus, we fit a linear function
to the Galactic trends of the abundance ratios, [X/Fe], of solar
analogs without detected planets \citep[see Fig.~3 in][]{gon10}, and
we subtract the value of these trends to each element abundance at the
metallicity of each star. The result is depicted in right panel of 
Fig.~\ref{fsa}. There one can see that the slopes of the previous 
linear fits to the data points tend to adopt values very close to 
zero. The scatter around the fits also decreases significantly. The
fact that we find very similar behaviour in the mean abundance pattern
of stars with and without planets, irrespective of the number of 
planets, and mass and orbital period of the planets, suggest that 
there is no clear signature of terrestrial planets in these mean 
patterns.

\subsection{Solar analogs with very low-mass planets}

\begin{figure}[!ht]
\centering
\includegraphics[width=4.7cm,angle=90]{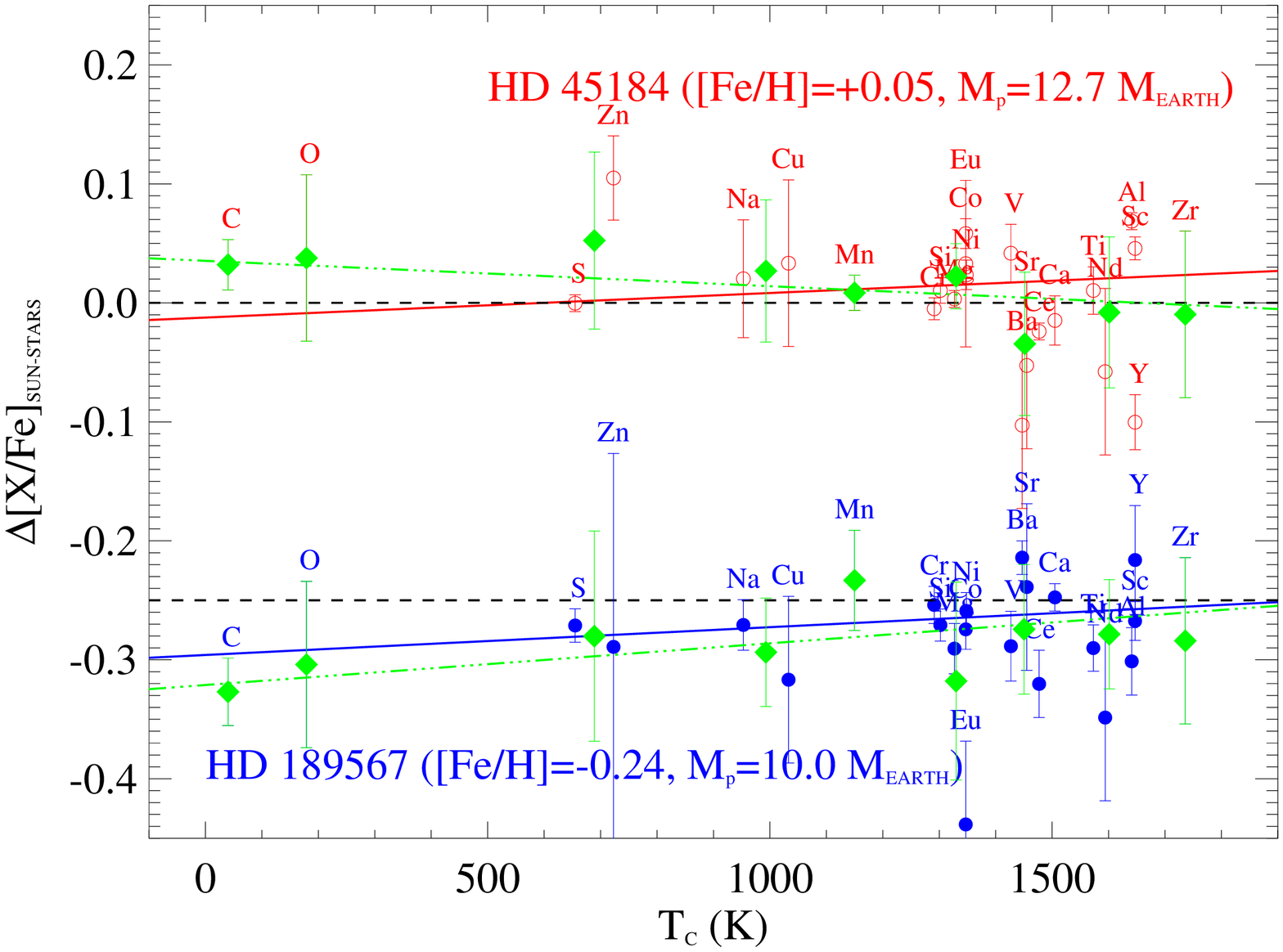}
\includegraphics[width=4.7cm,angle=90]{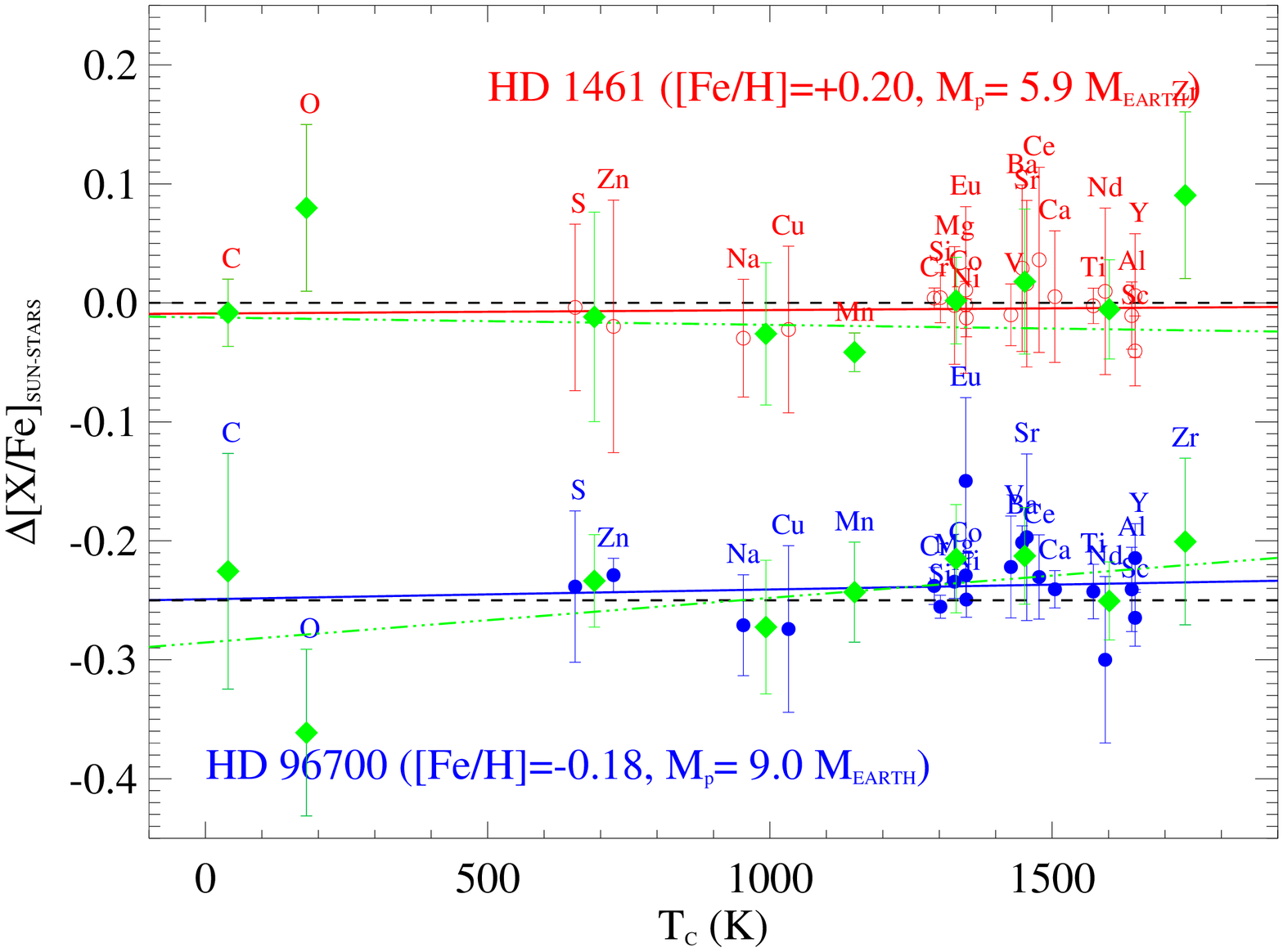}
\caption{\scriptsize{{\it Left panel:} Abundance differences, $\Delta {\rm [X/Fe]_{SUN-STARS}}$, 
between the Sun and two solar analogs hosting only one 
super-Earth-like planet each (circles).
Error bars are the uncertainties of the element abundance 
measurements, corresponding to the line-by-line scatter. 
Diamonds show the average abundances in bins of $\Delta T_C = 150$~K. 
Error bars are the standard deviation from the mean abundance of
the elements in each $T_C$ bin.
Linear fits to the data points (solid line) and to the mean data 
points (dashed-dotted line) weighted with the error bars are 
also displayed. 
An arbitrary shift of -0.25~dex has been applied to the abundances 
of one planet host, for the sake of clarity.
{\it Right panel:} Same as left panel but these two stars
harbour only two super-Earth-like planets each.}}  
\label{fse}
\end{figure}

Eight of the solar analogs host super-Earth-like planets and four 
stars harbour Neptune-like planets.  
We try to investigate the individual patterns of these 
stars (with respect to the Sun) to search for a 
signature of terrestrial planets. 
In Fig.~\ref{fse} we display the abundance pattern of four stars 
hosting super-Earth-like planets. We have removed the Galactic
chemical effects from the individual abundance ratios of these stars. 
The linear fits depicted as solid
lines give the same weight to each element abundance but there are
much more refractory elements than volatiles. Therefore, we think a
linear fit that attach the same importance to all condensation
temperature values would have more significance. Thus, we compute the
mean value of the element abundances in bins of $\Delta T_C = 150$~K,
using all the element abundances available in each $T_C$ bin. The
result is also shown in Fig.~\ref{fse}. The slope given by these mean fits
change sometimes with respect to the linear fit of all the element
abundance ratios taken separately. The solar analogs HD~45184 and 
HD~189567 host only one super-Earth-like planet each with a 
minimum mass $m_p \sin i \sim$~12.7 and 10.0~$M_\oplus$, respectively. 
The stars HD~1461 and HD~96700 harbour only two super-Earth-like 
planets each with $m_p \sin i \sim$~5.9 and 7.6~$M_\oplus$, and 
$m_p \sin i \sim$~9.0 and 12.7~$M_\oplus$, respectively. 
If one assumes that the
super-Earth-like planets mostly contain rocky material and therefore
are abundant in refractory elements, then according to the line of
reasoning in \citet{mel09}, these stars should exhibit mean 
abundance trends with positive slopes. This is not the case for 
the star HD~45184. The star HD~1461 shows a flat abundance pattern 
that it is also not expected since the amount of rocks in the two
super-Earth-like planets should be significantly greater than in the
solar system. The solar analogs HD~189567 and HD~96700 hosting one 
and two super-Earth-like planets, respectively, show increasing
abundance trends versus increasing condensation temperature. 
Although for these two stars we find the expected positive slopes,
the absolute value of this slope is larger for the star hosting only
one planet, which is {\it a priori} not expected.

The stars whose less massive planet is a Neptune-like planet, 
which we assume {\it ad hoc} to have a minimum mass in the range 
$m_p \sin i \sim$~14--30~$M_\oplus$, also exhibit different
behaviours. On the one hand, two of them show high (positive) 
values of the slopes of their abundances trends versus $T_C$. Hence 
this would mean that they should contain a substantial 
amount of rocky material forming low-mass planets, maybe pointing 
to these Neptune-like planets. On the other hand, the other two stars,
with possibly similar Neptune-like planets,
show negative slopes, which is not consistent with the previous line
of reasoning. 

\section{Conclusions}

We have carefully inspected the abundance patterns as a function of
condensation temperature of solar analogs trying to identify any 
clear signature of terrestrial planets. 
The result is not conclusive enough. The mean abundance ratios 
versus condensation of stars with and without detected planets 
exhibit similar behaviours which may indicate that if there is any 
signal, should be the same for all stars, irrespective of whether
these stars host planets or not. We note that most of the stars 
with planets are indeed stars with known giant planets. 

However, some stars harbour already detected low-mass planets. In
particular, in this sample of solar analogs, there are eight stars 
hosting super-Earth-like planets, and four with Neptune-like planets,
which present different volatile-to-refractory abundance ratios. 
One would expect that the amount of rocky material (and hence the
content of refractory elements) in low-mass planets is 
significant.
Only some of these stars show positive trends versus condensation 
temperature, which is what one would expect if these trends hide
information related to the presence of rocks in their planetary
systems. The other stars display flat or negative trends which is in
disagreement with the previous statement.
There might be a connection between the abundance patterns of stars,
in particular, solar analogs, and the presence of terrestrial planets
but this work reveals that there is still no clear evidence.

\acknowledgments

J.I.G.H. and G.I. acknowledge financial support from the 
Spanish Ministry project MICINN AYA2011-29060 and J.I.G.H. 
also from the Spanish Ministry of Science and Innovation 
(MICINN) under the 2009 Juan de la Cierva Programme. 
This work was also supported by the European Research
Council/European Community under the FP7/EC through a Starting 
Grant agreement number 239953, as well as by Funda\c{c}\~ao 
para a Ci\^encia e Tecnologia (FCT) in the form of grant 
reference PTDC/CTE-AST/098528/2008. 
N.C.S. would further like to thank FCT through program 
Ci\^encia 2007 funded by FCT/MCTES (Portugal) and POPH/FSE (EC). 
S.G.S. is supported by grants SFRH/BPD/70574/2010 
and SFRH/BPD/47611/2008 from FCT (Portugal), respectively.
E.D.M is supported by grant SFRH/BPD/76606/2011 from FCT (Portugal).
This research has made use of the SIMBAD database
operated at CDS, Strasbourg, France. 
This work has also made use of the IRAF facility, and the 
Encyclopaedia of extrasolar planets.

\bibliography{planetasp}

\end{document}